\documentclass[manuscript]{aastex}
\usepackage{graphicx}
\usepackage{epstopdf}

\shorttitle{Voids and Galaxy Number Counts} \shortauthors{Bucklein et al.}

\begin{document}


\title{The Effect of Nearby Voids on Galaxy Number Counts}


\author{Brian K. Bucklein}

\affil{Department of Computer Science, Math, \& Physics, Missouri Western State University, St. Joseph, MO 64507}

\and

\author{J. Ward Moody \& Eric G. Hintz}
\affil{Department of Physics and Astronomy, Brigham Young University, N283 ESC, Provo, Utah 84602}

\email{bbucklein@missouriwestern.edu, jmoody@physics.byu.edu, hintz@physics.byu.edu}


\begin{abstract}
The size, shape and degree of emptiness of void interiors sheds light on the details of galaxy formation.  A particularly interesting question is whether void interiors are completely empty or contain a dwarf population. However the nearby voids that are most conducive for dwarf searches have large angular diameters, on the order of a steradian, making it difficult to redshift-map a statistically significant portion of their volume to the magnitude limit of dwarf galaxies. As part of addressing this problem, we investigate here the usefulness of number counts in establishing the best locations to search inside nearby ($d <$ 300 Mpc) galaxy voids, utilizing Wolf plots of $\log(n < m)$ vs. $m$ as the basic diagnostic. To illustrate expected signatures, we consider the signature of three void profiles, ``cut out'', ``built up'', and ``universal profile'' carved into Monte-Carlo Schechter function models. We then investigate the signatures of voids in the Millennium Run dark matter simulation and the Sloan Digital Sky Survey. We find in all of these the evidence for cut-out and built-up voids is most discernible when the void diameter is similar to the distance to its center. However the density distribution of the universal profile that is characteristic of actual voids is essentially undetectable at any distance. A useful corollary of this finding is that galaxy counts are a reliable measure of survey completeness and stellar contamination even when sampling through significant voids.
\end{abstract}




\section{Introduction}
It is well-known that galaxies are assembled into a ``cosmic web'' of clusters, filaments, and sheets surrounding under-dense voids. The distribution of galaxies within it, particularly in the low density volumes, gives insight into galaxy formation and evolution and the role dark matter plays in it (e.g. \citealp{Benson03} and references therein). Lambda cold dark matter ($\Lambda$CDM) models predict the existence of many low-mass dark-matter halos in voids (e.g. \citealp{Dek86,Peb01,Hoff92,Tik09}). If galaxy formation has proceeded in these halos then voids should have a population of smaller galaxies in their interior (e.g. \citealp{Tik09}). But studies like \citet{Hoy05}, do not find them. This problem was termed the ``void phenomenon'' by \citet{Peb01}.

\citet{Tink09}, using an approach to galaxy biasing similar to that of \citet{Croton06}, found a good match between theory and observation for the low-density luminosity function (LMF), nearest neighbor statistics for dwarfs, and the void probability function of faint galaxies.  They predict that voids should be empty of dwarf galaxies fainter than seven magnitudes below $M^*$ or about $M_{r'} = -14$. They further predict that the drop-off in galaxy absolute magnitudes when transitioning from a filament to a void can be as steep as five magnitudes over $1 h^{-1}$ Mpc $(h = H_o/100)$.  Therefore the presence or absence of dwarf galaxies in voids and the abruptness of void boundaries differentiate between these two models.

\citet{Fos09} have compiled a list of nearby voids in the Updated Zwicky Catalog.  The closest of these are excellent places to search for dwarfs.  But their angular diameters are large, on the order of a steradian. Searching for dwarfs in their centers requires obtaining spectroscopic redshifts for a complete sample fainter than $r'$ of 20, a prohibitive task. For this reason surveys have concentrated on small fields or a subset of the galaxy population such as galaxies with emission (e.g.  \citealp{Sand84,Mood88,Kiss}). To help survey more efficiently we are revisiting the use of galaxy number counts to find the best possible places to search.

Galaxy number count (GNC) analysis is an accessible tool with a venerable history in clustering studies. Its many applications include galaxy luminosity evolution \citep{Metcalfe06, Bershady98}, mapping galactic extinction \citep{Fukugita04, Yasuda07}, mapping the extent of external galaxies \citep[e.g.][]{Ellis07}, delineating large-scale structure \citep{Frith06, Dolch05, Fukugita04}, exploring galaxy star-formation evolution \citep{Kong06}, mapping galaxy x-ray evolution \citep{Georgakakis06}, determining the AGN fraction \citep{Treister06}, and better understanding galaxy formation processes \citep{Berrier06, Lopez04}.

In this report we investigate the suitability of using GNC analysis to map the extent of emptiness of nearby galaxy voids. As a diagnostic, we utilize plots of $\log(n < m)$ vs. $m$, called ``Wolf plots'' after the work of \citet{Wolf23,Wolf24,Wolf26,Wolf32}, who used them to study stellar extinction. If space is uniformly filled with objects of an arbitrary but unchanging magnitude distribution, it is straight-forward to show that the slope of $\log(n < m)$ vs $m$ has a constant value of 0.6 in a complete magnitude-limited survey of objects that do not evolve with depth. A slope greater than 0.6 indicates a spatial overdensity in the survey volume, while a slope less than 0.6 indicates an underdensity or void. \citet{Koo86} has shown that evolutionary effects cause the slope to vary between 0.43 and 0.68 depending on depth and survey filter but the slope itself is reasonably constant over magnitude spans of interest here.

In this report we consider how voids can be understood using Wolf plots. We first look at their behavior in Monte-Carlo models of uniform galaxy distributions within which a number of different shaped voids were carved.  We then examine three voids from the Millennium Run simulation \citep{Springel05} and the semi-analytic galaxy catalog of \citet{Croton06}. Finally we examine a section of the Sloan Digital Sky Survey (SDSS) Data Release 7 \citep{Abazajian09} photometric data set.

\section{Applications to Possible Galaxy Distributions}
\subsection{The Uniform Distribution}
We began with simple void profiles to quantify how void distance, diameter, and edge density impact a Wolf plot. We assumed a galaxy luminosity function (LMF) that follows the standard \citet{Schechter76} function of
\begin{equation}
\phi(L)dL = \phi^{\ast}(L/L^{\ast})^{\alpha}\exp(-L/L^{\ast})d(L/L^{\ast})\label{eq:SchechterLum}
\end{equation}
where $\phi^{\ast}$ is the spatial density, $L^{\ast}$ is the turnover point of the LMF knee, and $\alpha$ is the slope of the faint end. For our models we expressed Equation~\ref{eq:SchechterLum} in terms of magnitudes as
\begin{equation}
\phi(M) dM = \frac{2}{5} \phi^{\ast} (\ln10)\left[10^{\frac{2}{5}\left(M^{\ast}-M\right)}\right]^{\alpha+1} \exp\left[-10^{\frac{2}{5}\left(M^{\ast}-M\right)}\right] dM \label{eq:SchechterFn}
\end{equation}
where $M$ is the absolute magnitude and $\phi(M) dM$ is the number of galaxies with magnitude between $M$ and $M + dM$ per Mpc$^{3}$.  All models were magnitude-limited with $\phi^{\ast}=0.90\times10^{-2}h^{3}$ Mpc$^{-3}$, $M^{\ast}=-20.73+5\log(h)$, and $\alpha=-1.23$.  These values are from the SDSS DR6 \citep{Montero09} and assume $h=1$. We limited the range of absolute magnitudes to $-24.0 < M_{r} < -14.0$. The sharp decrease of the Schechter function at bright magnitudes and the small volume sampled for the faint end guarantee that these limits do not exclude a significant number of galaxies.

The model first randomly assigned absolute magnitudes according to the Schechter function.  It then calculated the maximum distance $d$ at which a galaxy of that absolute magnitude would have an apparent magnitude brighter than the completeness limit $m_{lim}$ of the survey. A volume-weighted distance was randomly assigned to each galaxy within the maximum distance limit, ensuring that their distribution was uniform with depth. Finally an apparent magnitude was calculated from the distance and the absolute magnitude. Since we are considering the near ($d <$ 300 Mpc) universe, we did not apply higher-order correction such as $K$-corrections, or those for spatial curvature.

\begin{figure}
\epsscale{0.9}
\plotone{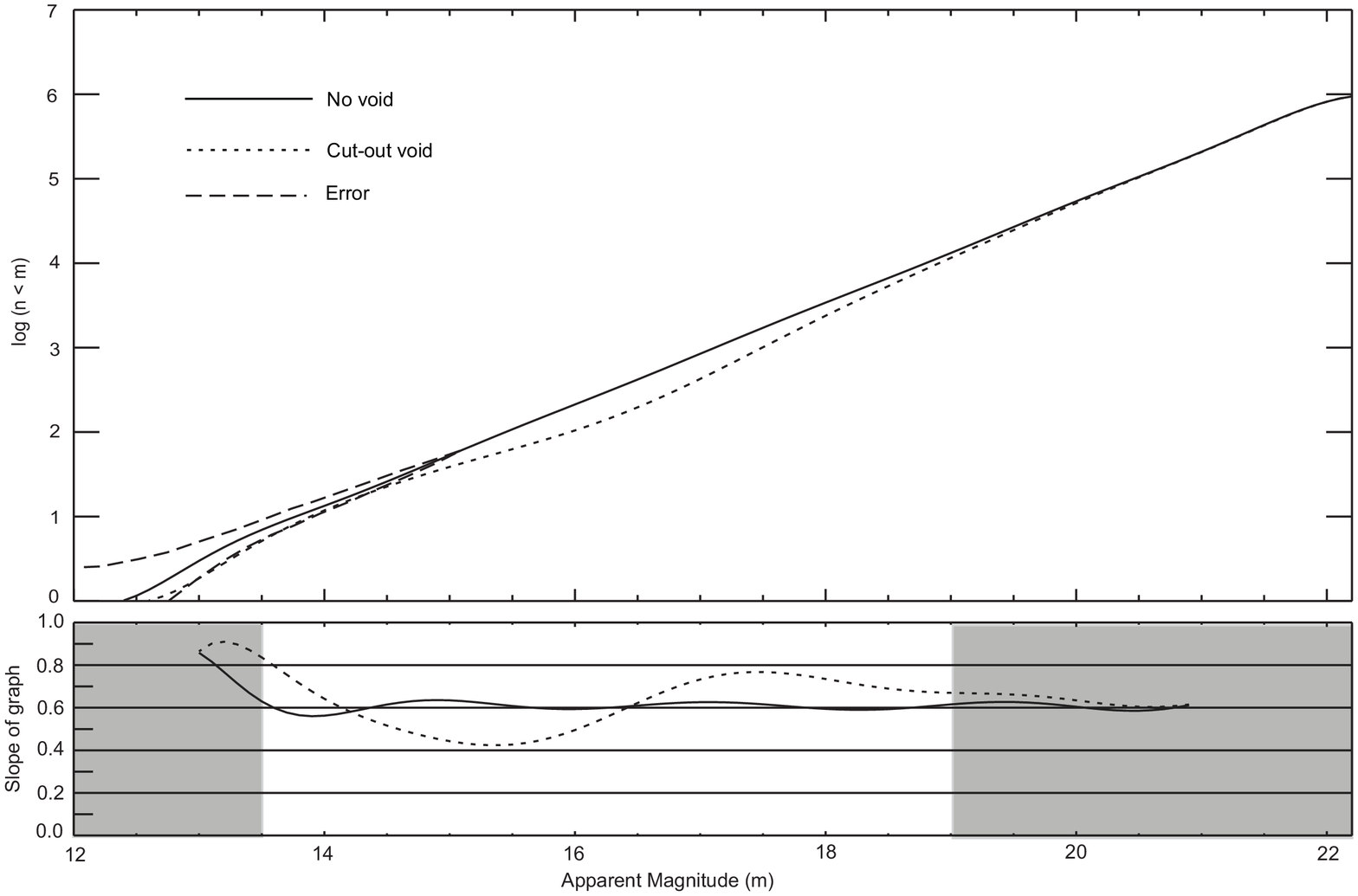}
\caption{Top: Wolf plots for the mean of 50 uniform galaxy distributions (solid line) and 50 uniform distributions with cut-out voids located at a distance of 250 Mpc having a width of 100 Mpc (dotted line). The one-sigma error envelope from Poisson statistics for the uniform distribution is outlined with a longer-dashed line. Bottom: The slope of a Legendre polynomial fit to the above data. Regions where the slope is dominated by small number statistics on the left and large numbers masking variation on the right are shaded in gray.}
\label{fig:WolfSOVoid}
\end{figure}

Figure~\ref{fig:WolfSOVoid} shows the averaged integrated and differential Wolf plots for 50 models of a distribution of galaxies that is $5^{\circ}$ in right ascension, and $1^{\circ}$ in declination having the SDSS limiting magnitude of $r = 22.2$. We applied a least-squares Legendre polynomial fit to the integrated data in the upper figure to generate the slope shown in the lower figure. The bright end of the data is dominated by small-number statistics which nearly always cause the slope to rise rapidly and overshoot the nominal 0.6 value, settling down to meaningful values only after an $m$ of 13.5.

\cite{Nadathur14} have determined using the watershed method of \cite{Neyrinck08} on both theoretical and SDSS data that the density distribution from void center to edge has a universal profile that is independent of scale size or tracer used to delineate structure. This remarkable find essentially answers the question of how the ``field'' galaxy density decreases from edge to center. We consider their find after first presenting two instructive distributions that bracket their profile; a ``cut-out'' void, and a ``built-up'' void. All models presented in the following subsections are the average of at 50 models run with the value of $\phi^{\ast}=0.90\times10^{-2}h^{3}$.

\subsection{Cut-Out Voids}\label{sub:CutOut}
Cut-out voids are completely empty holes in an otherwise uniform distribution.  They were created by randomly re-assigning the distances to galaxies that fell inside the void until they were outside the void boundaries. This preserved the value of $\phi^{\ast}$, produced a completely empty void with sharply defined edges, and created no clustering on the void front or back edges.

Figure~\ref{fig:WolfSOVoid} shows the Wolf plot for a 25 square degree sample of a cut-out void located at $d_{center} = 250$~Mpc with a width of 100~Mpc.  By $m=14.0$ the slope has settled at about 0.6.  It dips down to 0.4 then increases back to $0.6$ at about $m=16.40$, which is close to the apparent magnitude of an $M^{\ast}$ galaxy located at the void center. The slope then plays ``catch-up'' by overshooting 0.6 before settling back down to this value at the faint end. This pattern is true of all cut-out voids regardless of width or distance: there is a decreasing slope followed by an increasing slope with the cross-over point being close to the apparent magnitude of an $M^{\ast}$ galaxy centered in the void.

Wolf plots have large fluctuations from small-number statistics on the bright end.  On the faint end large number statistics function as noise masking variations from voids. To better highlight the span of greatest information, we shade those regions that were least sensitive to the presence of a void.  For most plots the span of $13.5 \leq m \leq 19.0$ is the most relevant segment.

\subsection{Built-Up Voids}
The built-up void models again started with a random distribution of galaxies.  This time galaxies within the void were swept onto the front or back edge, whichever was closest. The location within the edge was determined by a probability function of the form $\cos^{2}(x)$ such that the center of the distribution was at the edge of the void, falling to zero by the time $x = void\_edge \pm \frac{1}{2} edge\_width$.  Voids produced this way have galaxies within them, depending on the edge width specified. We chose a $\cos^{2}(x)$ relation over other choices such as a Gaussian to better truncate the edge distribution at specific widths.

\begin{figure}
\epsscale{0.9}
\plotone{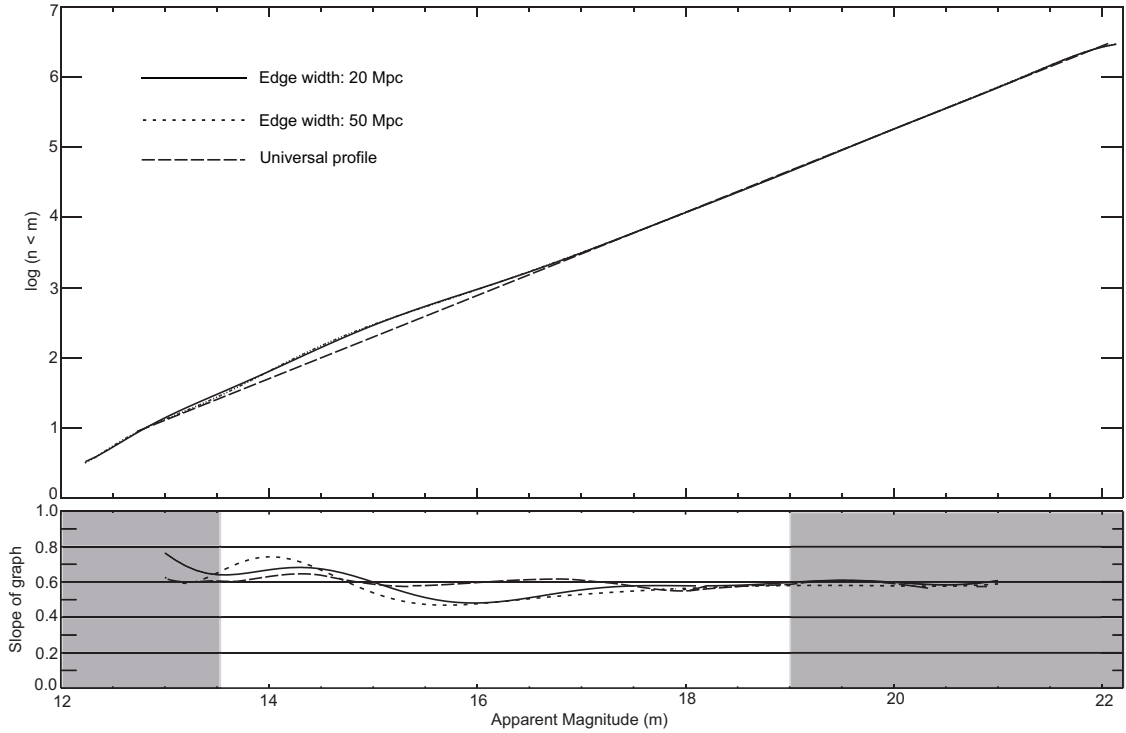}
\caption{Top: Wolf plots of the mean of 50 uniform galaxy distributions with built-up voids located at a distance of 250 Mpc having a width of 100 Mpc and an edge width of 20 Mpc (solid line), 50 Mpc (dotted line) and the universal profile of \cite{Nadathur14} (dashed line). The built-up voids are remarkably similar despite the edge differences. The universal void profile is a straight line to within the errors of the plot.}
\label{fig:WolfBUCombined}
\end{figure}

Figure~\ref{fig:WolfBUCombined} shows the Wolf plots for voids centered at $d = 250$ Mpc and widths of $100$ Mpc with edge widths of $20$ Mpc and $50$ Mpc respectively.  As can be seen, the chosen value of the edge width does not create a significant difference.  The build-up on the front edge of the void increases the slope above $0.6$.  The slope then decreases to below the expected value, and eventually returns to $0.6$.  This is characteristic behavior for all built-up voids, regardless of width or distance.  The point where the slope crosses over from above to below 0.6 is always brighter than for the cut-out voids, occurring at $m=15.00$ and $m=14.80$ for an edge width of 20 Mpc and 50 Mpc in the cases presented here. In general, the wider the width, the brighter the cross-over point.

\subsection{Universal Profile}

To create voids having the universal profile of \cite{Nadathur14} we first started with a uniform distribution having a density of $2\phi^{\ast}$.  Within this distribution a void was created at a distance of 250 Mpc and a width of 100 Mpc as before.  To create the void, each galaxy was flagged and kept with a probability given by equation 9 from their paper. The probability of acceptance was scaled so that the overall probability of rejection was 0.5, making the resultant density $\phi^{\ast}$.

Figure~\ref{fig:WolfBUCombined} shows the result for the average of 50 models. There is no deviation from a straight line to within the errors. This is not totally surprising since the universal profile has a greater density on the edge then tapers off into a mostly empty void. These two aspects affect Wolf plots oppositely, as shown above, to where there is essentially no net signature. Even in sampling the equivalent of a 50 square degree void projected area, there is no statistically significant void signature.

\subsection{Resolution}\label{sub:Detection with Distance}
\begin{figure}
\epsscale{0.95}
\plotone{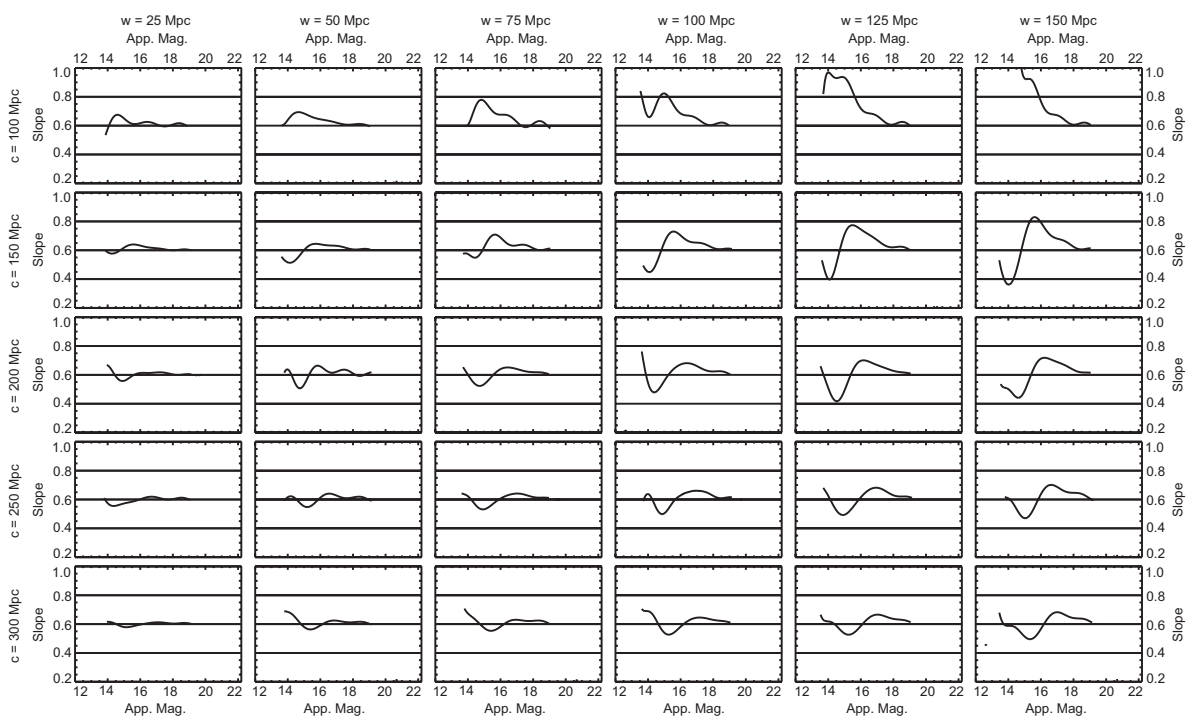}
\caption{A grid of Wolf plot slopes showing how the signature of a cut-out void changes when going from a center distance of 100 Mpc to 300 Mpc and a width of 25 Mpc to 150 Mpc. Distance increases down each column while the width increases across each row.  Each plot is the average of 50 models.}
\label{fig:ResolutionGrid}
\end{figure}

The previous sections considered void profiles for a given width and distance, To illustrate how the detectability of a void changes as size and distance change, Figure~\ref{fig:ResolutionGrid} presents a grid of Wolf plots for cut-out voids. We use cut-out voids because they have an easily recognized signature and the point of the figure is to illustrate \emph{trends} in signature. The trend for built-up voids is the same so we do not present it as well.

The width of the void in this figure increases left to right from 50 Mpc to 150 Mpc in 25 Mpc increments, while the distance to the center of the void increases top to bottom from 100 Mpc to 300 Mpc in 50 Mpc increments.  As before, all of the voids in this figure are the average of 50 models.  We refer to each void as cXXXwYYY, where XXX specifies the distance to the center and YYY indicates the width.

The importance of enough counts to anchor the front side is illustrated by the first row where the slope from a magnitude of 13.5 to 15.0 climbs from around 0.6 to over 1.0. This is because as the width increases across the row, the front side thins until it is essentially gone by c100w150. The ``dip-bump'' signature of a cut-out void is visible in the first three frames with the overall slope rising as the back side begins to dominate. By c100w150 there is a sudden increase in counts causing a slope spike at the fainter magnitude of the back side followed by a return to a uniform distribution.

The effect from increasing width is seen best in rows two and three. In c150w025, the dip-bump profile is apparent but not at a significant level. By c150w100 it is readily apparent, growing into a classical signature by c150w150.  The same trend is visible on the lower rows with a decreased amplitude as expected. This illustrates that in general, the signature of a void of any profile shows up best for voids whose diameter is about the same value as their distance from us.

The effect from increasing distance is illustrated by looking down a single column.  Consider the graphs from c150w150 to c300w150.  Going down the column the amplitude decreases  and the slope cross-over point gets fainter, both as expected. In c150w150, the cross-over point occurs at $m \sim 14.75$.  By c300w150 it is at $m \sim 16.25$.  As previously mentioned these values correlate with the apparent magnitude of an $M^{\ast}$ galaxy located at the void center.

\section{Applications to Data}
\subsection{The Millennium Run} \label{sec:GNCMillRun}
The above section presented the signatures that might be seen in a Wolf plot. We next formed Wolf plots of voids in the Millennium Run simulation of \citet{Springel05} to see if there were any similarities. The Millennium Run is based on the 2-degree Field Galaxy Redshift Survey (2dFGRS) described in \citet{Colless01}, \citet{Folkes99}, and \citet{Cole05}, and uses the first-year Wilkinson Microwave Anisotropy Probe (WMAP) data in \citet{Bennett03} and \citet{Spergel03} to infer primordial density fluctuations. They traced more than $10^{10}$ dark matter particles, each with a mass of $8.6\times10^{8} h^{-1}$ M$_{\odot}$ in a periodic cubical region $500 h^{-1}$ Mpc on a side then populated dark matter sites with galaxies as outlined in \citet{Springel05} and \citet{Croton06}.  This resulted in $\sim9.5$ million galaxies with $M_{r} \le -16$.

We created slices through the Millennium Run that were 10~Mpc thick in the z-coordinate, with each slice incrementing z by 5~Mpc. We visually searched through these for promising voids within 250~Mpc of the origin.  We formed Wolf plots of these and found their signatures to most resemble cut-out voids but the scatter in slope profiles was large.  

\begin{figure}
\epsscale{0.9}
\plotone{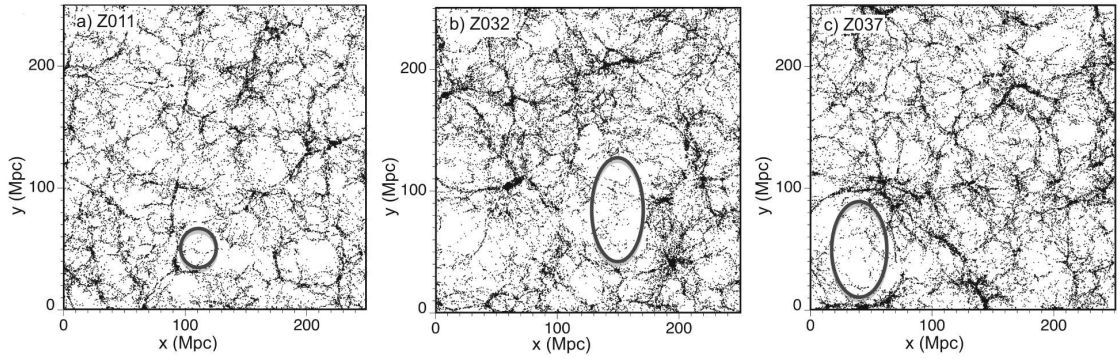}
\caption{Slices of thickness 10 Mpc through the Millennium Run data set showing the location of voids a) Z011 (distance = 135 Mpc, width = 40 Mpc) b) Z032 (distance = 240 Mpc, width = 60 Mpc), and c) Z037 (distance = 200 Mpc, width = 100 Mpc). The observer is located at the 0,0,0 origin.}
\label{fig:MRVoidsCombined}
\end{figure}

\begin{figure}
\epsscale{0.9}
\plotone{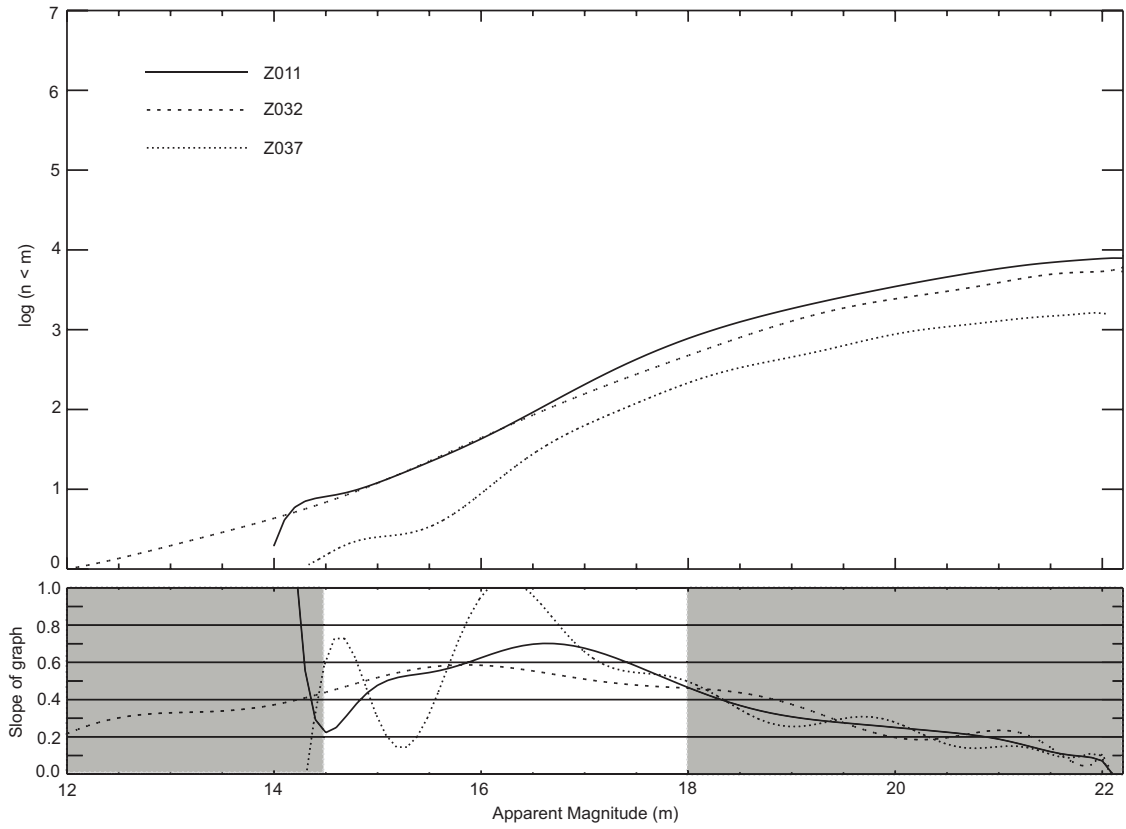}
\caption{Wolf plots for voids Z011, Z032, and Z037 given in Figure~\ref{fig:MRVoidsCombined}.  The signature of a cut-out void is apparent in voids Z011 and Z037.}
\label{fig:WolfMRVoidsCombined}
\end{figure}

To illustrate typical results we present three of them called Z011, Z032, and Z037.  These designations refer to the slice in which the center of the void is located (i.e. Z011 is  the slice 110 Mpc from the origin in the z direction). The voids are shown in their respective slices in Figure~\ref{fig:MRVoidsCombined} and the Wolf plots are shown in Figure~\ref{fig:WolfMRVoidsCombined}.

Void Z011 is located approximately 135 Mpc from the origin, which we took to be the observer's location. Its diameter is about 40 Mpc. We calculated apparent magnitudes for a five square degree circle directed at the void heart to form the Wolf plot of Figure~\ref{fig:WolfMRVoidsCombined}. A cut-out void pattern is visible having a dip-bump amplitude that is consistent with our models for similar distances and diameters (e.g. c150w050). The ascending slope cross-over occurs at about $m=16.0$, fainter than expected from the previous simulations but within the variations of our models. Because this is a volume-limited sample, the slope in this and the other voids bends over and tails off fainter than an $m$ of 17.

Void Z032 is located at a distance of approximately 240 Mpc with a diameter along the line of sight of about 60 Mpc. The Wolf plot slope for a five degree by one degree line-of-sight from the origin to the void center never does reach 0.6. A field of view twice as large made no difference. Despite the appearance in Figure~\ref{fig:MRVoidsCombined}, there is little population in the foreground before encountering the void which affects the results.

The last void we considered, Z037, has a center located midway between the distances of the previous two voids, at approximately 200 Mpc, and a diameter of about 100 Mpc.  There are few galaxies on the front side of the void, raising the same concerns as with the other voids. However, the back side of void Z037 is well delineated, and the front side has just enough population to anchor the slope near 0.6, allowing the classic cut-out void signature to emerge. It falls and rises by more than $\pm 0.4$, which is similar to void c150w150 in Figure~\ref{fig:ResolutionGrid}.  Since the void is centered at about 200 Mpc, this suggests a void diameter on the order of 200 Mpc.

As stated above, these three voids illustrate the general properties of voids in the Millennium Run.  Most Wolf plot profiles were dominated by small number statistics on the front side. Five square degrees is not enough volume to get more than a several dozen galaxies even for more distant voids. When the front side is not well anchored the Wolf plot slope profile resembles a cut-out void even if the profile is built-up or universal. Even so, a bit to our surprise, the signature of voids was visible to a degree as great as was found in our simple Schechter function models.

\subsection{The Sloan Digital Sky Survey}
we next applied Wolf plots to Sloan Digital Sky Survey (SDSS) photometric data from DR7 (\citealp{Abazajian09}).  These data are complete to $m_{r} = 22.2$. \citep{Lup01} has shown that galaxies are correctly identified in this data set approximately 95\% of the time. 

\begin{figure}
\epsscale{0.9}
\plotone{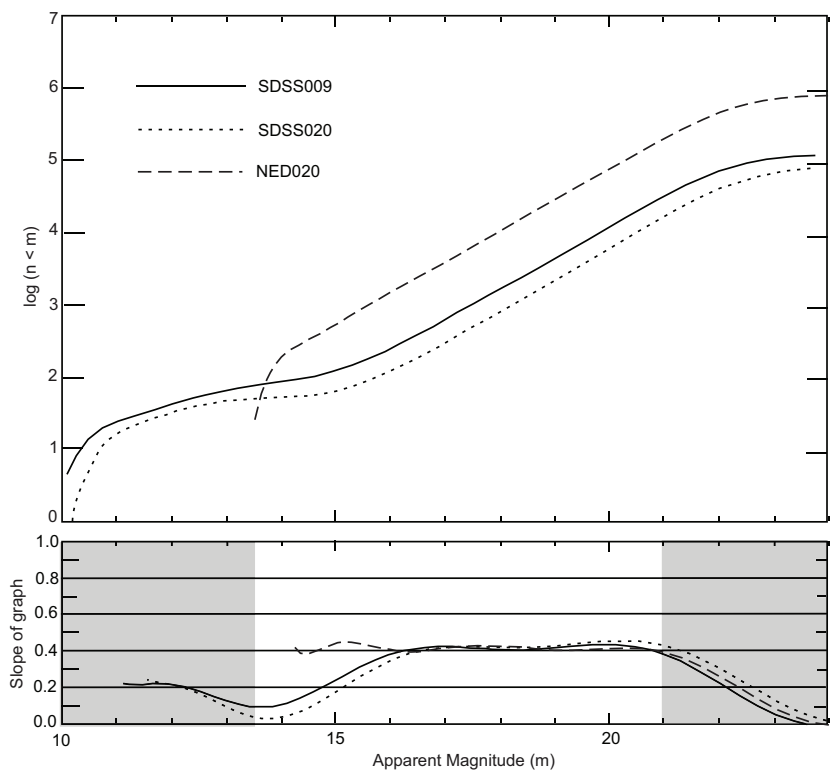}
\caption{Wolf plots of SDSS data for fields a) SDSS009 (RA = 10h 20m, Dec = $3.5^{\circ}$), b) SDSS020 (RA = 14h 30m, Dec = $43^{\circ}$), and c) NED020 (RA = 14h 30m, Dec = $43^{\circ}$).  The SDSS fields are 5 square degrees and the NED field is 50 square degrees.  The large plateau in the SDSS fields is from stars and galaxy pieces being misidentified as galaxies. Field NED020 incorporates the same area as SDSS020 but with mis-identified objects removed.  No signature of the Bo\"{o}tes void is discernible in this field.}
\label{fig:SDSSWolfCombined}
\end{figure}

Figure~\ref{fig:SDSSWolfCombined} shows the results for two 5 degree by 1 degree lines-of-sight centered at RA = 10h 20m, Ded = 3.5$^{\circ}$ and 14h 30m 43$^{\circ}$ respectively. Conspicuous in both plots is a large plateau brighter than $m_{r} = 15.0$ that is similar to the signature of a nearby cut-out void. All plots of SDSS data show this same signature with the length of the plateau decreasing with higher galactic latitude. We confirmed by manual inspection that the plateau is from the small percentage of stars that are misidentified as galaxies. These mistaken identifications are more prevalent for objects brighter than $m_{r} = 15.0$, falsely inflating the beginning of the curve.

With the assistance of personnel at the NASA/IPAC Extragalactic Database (NED), we obtained from NED a $10^{\circ} \times 5^{\circ}$ segment of SDSS data centered on RA = 14h 30m, Dec  $217.0^{\circ}$ right ascension and $43.0^{\circ}$ declination).  These data were reprocessed to remove pieces of brighter galaxies and merged stars falsely identified as galaxies. As Figure~\ref{fig:SDSSWolfCombined} shows, the plateau is completely removed.

The NED sample is in the direction of the Bo\"{o}tes void which is centered at 250 Mpc with a diameter of about 80 Mpc.  That no void signature is evident in the NED processed data illustrates how insensitive Wolf plots are to detect this void. We applied this technique to several known voids, all with the same result. 

\section{Conclusion}
Both simple and more complicated simulations suggest that Wolf plots are capable of revealing voids in the universe closer than 300 Mpc, if they have a cut-out type density profile. That some may have this type of profile is suggested by Millennium Run data. This suggests that it is possible to find corridors of emptiness by making a grid of Wolf plots in the area of interest and looking for the signature of a cut-out void.

But getting statistics large enough for a well-formed Wolf plot requires sampling large areas of the sky.  Features smaller than the areal extent are quickly averaged out to where they cannot be detected. On top of this, the universal profile of \cite{Nadathur14} is nicely masked in Wolf plots. Therefore Wolf plots of the entire galaxy population, as was done here, do not appear to be an effective way to locate or map typical voids.  The consolation is that they are a reasonable way to check for survey contamination and completeness.

A central question of this study is whether Wolf plots are capable of revealing the best places to look for dwarf galaxies inside voids.  Unfortunately the lack of a signature from the universal profile together with the faintness of dwarfs mitigates against this. Wolf plots in particular and GNC analysis in general average across too great a population spread too far in space to reveal a signature by themselves. We conclude that spectroscopic analysis of individual candidates chosen through other means is still a superior approach.

\acknowledgments We thank Joe Mazarella and Rick Ebert of IPAC for many useful conversations and assistance.  We gratefully acknowledge the Missouri Western State University Department of Computer Science, Mathematics, and Physics and the Brigham Young University Department of Physics and Astronomy for their support.





\clearpage



\clearpage


\begin{thebibliography}{}
\bibitem[Abazajian et al.(2009)]{Abazajian09} Abazajian et al. 2009, \apjs, 182, 543
\bibitem[Bennett et al.(2003)]{Bennett03} Bennett, C. L., et al. 2003, \apjs, 148, 1
\bibitem[Benson et al.(2003)]{Benson03} Benson, A. J., Hoyle, F., Torres, F., and Vogeley, M. S. 2003, \mnras, 340, 160
\bibitem[Berrier et al.(2006)]{Berrier06} Berrier, J. C., Bullock, J. S., Barton, E. J., Guenther, H. D., Zentner, A. R. \& Wechsler, R. H. 2006, \apj, 652, 56
\bibitem[Bershady et al.(1998)]{Bershady98} Bershady, M. A., Lowenthal, J. D. \& Koo, D. C. 1998, \apj, 505, 50
\bibitem[Cole et al.(2005)]{Cole05} Cole, S., et al. 2005, \mnras, 362, 505
\bibitem[Colless et al.(2001)]{Colless01} Colless, M., et al. 2001, \mnras, 328, 1039
\bibitem[Croton et al.(2006)]{Croton06} Croton, D. J., Springel, V., White, S. D. M., De Lucia, G., Frenk, C. S., Gao, L., Jenkins, A., Kauffmann, G., Navarro, J. F., Yoshida, N. 2006, \mnras, 365, 11
\bibitem[Dekel \& Silk(1986)]{Dek86} Dekel, A. \& Silk, J. 1986, \apj, 303, 39
\bibitem[Dolch \& Ferguson(2005)]{Dolch05} Dolch, T. \& Ferguson, H. C. 2005, \baas, 37, 1193
\bibitem[Draper et al.(2016)]{Drap16} Draper, C., Moody, J. W., McNeil, S. M., and Joner, M. D. 2016, in preparation
\bibitem[Ellis \& Bland-Hawthorn(2007)]{Ellis07} Ellis, S. C. \& Bland-Hawthorn, J. 2007, \mnras, 337, 815
\bibitem[Folkes et al.(1999)]{Folkes99} Folkes, S., et al. 1999, \mnras, 308, 459
\bibitem[Foster \& Nelson(2009)]{Fos09} Foster, C. \& Nelson, L. A. 2009, \apj, 699, 1252
\bibitem[Frith et al.(2006)]{Frith06} Frith, W. J., Metcalfe, N., \& Shanks, T. 2006, \mnras, 371, 1601
\bibitem[Fukugita et al.(2004)]{Fukugita04} Fukugita, M., Yasuda, N., Brinkmann, J., Gunn, J. E., Ivezi\'{c}, {\v{Z}}., Knapp, G. R., Lupton, R. \& Schneider, D. P. 2004, \aj, 127, 3155
\bibitem[Georgakakis et al.(2006)]{Georgakakis06} Georgakakis, A., Georgantopoulos, I., Akylas, A., Zezas, A. \& Tzanavaris, P. 2006, \apj, 641, 101
\bibitem[Hoffman, Silk, \& Wyse(1992)]{Hoff92} Hoffman, Y., Silk, J., \& Wyse, R. F. G. 1992, \apj, 388, L13
\bibitem[Hoyle et al.(2005)]{Hoy05} Hoyle, F., Rojas, R. R., Vogeley, M. S., and Brinkmann, J. 2005, \apj, 620, 618
\bibitem[Huchra et al.(1988)]{Huchra88} Huchra, J. P., Geller, M. J., de Lapparent, V., \& Burg, R. 1988, IAU Symp., 130, 105
\bibitem[Jangren et al.(2005)]{Kiss} Jangren, A., Salzer, J. J., Sarajedini, V. L., Gronwall, C., Werk, J. K.,  Chomiuk, L. B., Moody, J. W., \& Boroson, T. 2005, \aj, 130, 2571
\bibitem[Kirshner et al.(1983)]{Kirshner83} Kirshner, R. P., Oemler, A., Jr., Schechter, P. L., \& Shectman, S. A. 1983, \aj, 88, 1285
\bibitem[Kong et al.(2006)]{Kong06} Kong, X., et al. 2006, \apj, 638, 72
\bibitem[Koo(1986)]{Koo86} Koo, D. C. 1986, \apj, 311, 651
\bibitem[L\'{o}pez-Corredoira \& Betancort-Rijo(2004)]{Lopez04} L\'{o}pez-Corredoira, M. and Betancort-Rijo, J.~E. 2004, \aap, 416, 1
\bibitem[Lupton et al.(2001)]{Lup01} Lupton, R., Gunn, J. E., Ivezi\'{c}, Z., Knapp, G. R., \& Kent, S. 2001 in ASP Conf. Ser. 238, Astronomical Data Analysis and Software Systems (ADASS) X, ed. F. R. Harnden Jr., F. A. Primini, and H. E. Payne (San Francisco: ASP) 269
\bibitem[Metcalfe et al.(2006)]{Metcalfe06} Metcalfe, N., Shanks, T., Weilbacher, P. M., McCracken, H. J., Fong, R. \& Thompson, D. 2006, \mnras, 370, 1257
\bibitem[Montero-Dorta \& Prada(2009)]{Montero09} Montero-Dorta, A. D. \& Prada, F. 2009, \mnras, 399, 1106
\bibitem[Moody(1988)]{Mood88} Moody, J.W., 1988, \pasp, 100, 1351
\bibitem[Nadathur et al.(2014)]{Nadathur14} Nadathur, S., Hotchkiss, S., Diego, J. M., Iliev, I. T., Gottl\"{o}ber, S., Watson, W. A., Yepes, G. 2014 \mnras in press, http://arxiv.org/abs/1407.1295
\bibitem[Neyrinck(2008)]{Neyrinck08} Nyerinck, M. C. 2008 \mnras 386, 2101
\bibitem[Peebles(2001)]{Peb01} Peebles, P. J. E. 2001, \apj, 557, 495
\bibitem[Sanduleak \& Pesch(1984)]{Sand84}Sanduleak, N. \& Pesch, P. 1984, \apjs, 55, 517
\bibitem[Schechter(1976)]{Schechter76} Schechter, P., 1976 \apj, 203, 297
\bibitem[Spergel et al.(2003)]{Spergel03} Spergel D. N., et al. 2003, \apjs, 148, 175
\bibitem[Springel et al.(2005)]{Springel05} Springel, V., et al. 2005, \nat, 435, 629
\bibitem[Tikhonov \& Klypin(2009)]{Tik09} Tikhonov, A. V. \& Klypin, A. 2009, \mnras, 395, 1915
\bibitem[Tinker \& Conroy(2009)]{Tink09} Tinker, J. L. \& Conroy, C. 2009, \apj, 691, 633
\bibitem[Treister et al.(2006)]{Treister06} Treister, E., et al. 2006, \apj, 640, 603
\bibitem[Wolf(1923)]{Wolf23} Wolf, M. 1923, AN, 219, 109
\bibitem[Wolf(1924)]{Wolf24} Wolf, M. 1924, AN, 223, 89
\bibitem[Wolf(1926)]{Wolf26} Wolf, M. 1926, AN, 229, 1
\bibitem[Wolf(1932)]{Wolf32} Wolf, M. 1932, ViHei, 8, 65
\bibitem[Yasuda et al.(2007)]{Yasuda07} Yasuda, N., Fukugita, M. \& Schneider, D. P. 2007, \aj, 134, 698


\end{thebibliography}
\end{document}